\documentclass[twocolumn,showpacs,preprintnumbers,amsmath,amssymb,prl]{revtex4}

\usepackage{graphicx}
\usepackage{dcolumn}
\usepackage{bm}

\newcommand{\ipb}{\ensuremath{\mathrm{pb^{-1}}}}

\newcommand{\TeV}{\ensuremath{\mathrm{Te\kern -0.1em V}}}
\newcommand{\TeVc}{\ensuremath{\mathrm{Te\kern -0.1em V\!/}c}}
\newcommand{\TeVcc}{\ensuremath{\mathrm{Te\kern -0.1em V\!/}c^2}}
\newcommand{\GeV}{\ensuremath{\mathrm{Ge\kern -0.1em V}}}
\newcommand{\GeVc}{\ensuremath{\mathrm{Ge\kern -0.1em V\!/}c}}
\newcommand{\GeVcc}{\ensuremath{\mathrm{Ge\kern -0.1em V\!/}c^2}}
\newcommand{\MeV}{\ensuremath{\mathrm{Me\kern -0.1em V}}}
\newcommand{\MeVc}{\ensuremath{\mathrm{Me\kern -0.1em V\!/}c}}
\newcommand{\MeVcc}{\ensuremath{\mathrm{Me\kern -0.1em V\!/}c^2}}

\newcommand{\rUP}[1]{\ignorespaces $^{#1}$}

\font\eightit=cmti8                                                                                

\begin{document}

\title{
{\raggedleft\small
\vspace*{-1.5cm}
FERMILAB-PUB-03/393-E\\
(Subm. Phys. Rev. Lett.)\\
}\vspace*{0.9cm}
Observation of the Narrow State $X(3872)\rightarrow J/\psi\pi^{+}\pi^{-}$\\
 in $\bar{p}p$ Collisions at $\sqrt{s}=1.96~\TeV$}

\author{ 
\hfilneg
\protect\begin{sloppypar}
\protect\noindent
D.~Acosta,\rUP {14} T.~Affolder,\rUP 7 M.H.~Ahn,\rUP {25} T.~Akimoto,\rUP {52}                            
M.G.~Albrow,\rUP {13} D.~Ambrose,\rUP {41} S.~Amerio,\rUP {40}   			  		   
D.~Amidei,\rUP {31} A.~Anastassov,\rUP {48} K.~Anikeev,\rUP {29}
% ...
\protect\end{sloppypar}	
\protect\vskip .026in				
\protect\begin{center}	
(CDF II Collaboration)										   
\protect\end{center}	
\protect\vskip .026in		
\protect\begin{center}	
\rUP 1  {\eightit Institute of Physics, Academia Sinica, Taipei, Taiwan 11529, 			   
Republic of China} \\										   
\rUP 2  {\eightit Argonne National Laboratory, Argonne, Illinois 60439} \\			   
\rUP 3  {\eightit Istituto Nazionale di Fisica Nucleare, University of Bologna,			   
I-40127 Bologna, Italy} \\									   
\rUP 4  {\eightit Brandeis University, Waltham, Massachusetts 02254} \\				   
\rUP 5  {\eightit University of California at Davis, Davis, California  95616} \\			   
\rUP 6  {\eightit University of California at Los Angeles, Los 					   
Angeles, California  90024} \\ 									   
\rUP 7  {\eightit University of California at Santa Barbara, Santa Barbara, California 		   
93106} \\ 											   
\rUP 8 {\eightit Instituto de Fisica de Cantabria, CSIC-University of Cantabria, 			   
39005 Santander, Spain} \\									   
\rUP 9  {\eightit Carnegie Mellon University, Pittsburgh, Pennsylvania  15213} \\			   
\rUP {10} {\eightit Enrico Fermi Institute, University of Chicago, Chicago, 			   
Illinois 60637} \\										   
\rUP {11}  {\eightit Joint Institute for Nuclear Research, RU-141980 Dubna, Russia}		   
\\												   
\rUP {12} {\eightit Duke University, Durham, North Carolina  27708} \\				   
\rUP {13} {\eightit Fermi National Accelerator Laboratory, Batavia, Illinois 			   
60510} \\											   
\rUP {14} {\eightit University of Florida, Gainesville, Florida  32611} \\			   
\rUP {15} {\eightit Laboratori Nazionali di Frascati, Istituto Nazionale di Fisica		   
               Nucleare, I-00044 Frascati, Italy} \\						   
\rUP {16} {\eightit University of Geneva, CH-1211 Geneva 4, Switzerland} \\			   
\rUP {17} {\eightit Glasgow University, Glasgow G12 8QQ, United Kingdom}\\			   
\rUP {18} {\eightit Harvard University, Cambridge, Massachusetts 02138} \\			   
\rUP {19} {\eightit The Helsinki Group: Helsinki Institute of Physics; and Division		   
of High Energy Physics, Department of Physical Sciences, University				   
of Helsinki, FIN-00014 Helsinki, Finland}\\							   
\rUP {20} {\eightit Hiroshima University, Higashi-Hiroshima 724, Japan} \\			   
\rUP {21} {\eightit University of Illinois, Urbana, Illinois 61801} \\				   
\rUP {22} {\eightit The Johns Hopkins University, Baltimore, Maryland 21218} \\			   
\rUP {23} {\eightit Institut f\"{u}r Experimentelle Kernphysik, 					   
Universit\"{a}t Karlsruhe, 76128 Karlsruhe, Germany} \\						   
\rUP {24} {\eightit High Energy Accelerator Research Organization (KEK), Tsukuba, 		   
Ibaraki 305, Japan} \\										   
\rUP {25} {\eightit Center for High Energy Physics: Kyungpook National				   
University, Taegu 702-701; Seoul National University, Seoul 151-742; and			   
SungKyunKwan University, Suwon 440-746; Korea} \\						   
\rUP {26} {\eightit Ernest Orlando Lawrence Berkeley National Laboratory, 			   
Berkeley, California 94720} \\									   
\rUP {27} {\eightit University of Liverpool, Liverpool L69 7ZE, United Kingdom} \\		   
\rUP {28} {\eightit University College London, London WC1E 6BT, United Kingdom} \\		   
\rUP {29} {\eightit Massachusetts Institute of Technology, Cambridge,				   
Massachusetts  02139} \\   									   
\rUP {30} {\eightit Institute of Particle Physics, McGill University, 				   
Montr\'{e}al, Canada H3A~2T8; and University of Toronto, Toronto, Canada			   
M5S~1A7}\\											   
\rUP {31} {\eightit University of Michigan, Ann Arbor, Michigan 48109} \\				   
\rUP {32} {\eightit Michigan State University, East Lansing, Michigan  48824} \\
\rUP {33} {\eightit Institution for Theoretical and Experimental Physics, ITEP,			   
Moscow 117259, Russia} \\									   
\rUP {34} {\eightit University of New Mexico, Albuquerque, New Mexico 87131} \\			   
\rUP {35} {\eightit Northwestern University, Evanston, Illinois  60208} \\			   
\rUP {36} {\eightit The Ohio State University, Columbus, Ohio  43210} \\  			   
\rUP {37} {\eightit Okayama University, Okayama 700-8530, Japan}\\  				   
\rUP {38} {\eightit Osaka City University, Osaka 588, Japan} \\					   
\rUP {39} {\eightit University of Oxford, Oxford OX1 3RH, United Kingdom} \\			   
\rUP {40} {\eightit Universit\'{a} di Padova, Istituto Nazionale di Fisica 			   
          Nucleare, Sezione di Padova-Trento, I-35131 Padova, Italy} \\				   
\rUP {41} {\eightit University of Pennsylvania, Philadelphia, 					   
        Pennsylvania 19104} \\   								   
\rUP {42} {\eightit Istituto Nazionale di Fisica Nucleare, University and Scuola			   
               Normale Superiore of Pisa, I-56100 Pisa, Italy} \\				   
\rUP {43} {\eightit University of Pittsburgh, Pittsburgh, Pennsylvania 15260} \\			   
\rUP {44} {\eightit Purdue University, West Lafayette, Indiana 47907} \\				   
\rUP {45} {\eightit University of Rochester, Rochester, New York 14627} \\			   
\rUP {46} {\eightit The Rockefeller University, New York, New York 10021} \\			   
\rUP {47} {\eightit Instituto Nazionale de Fisica Nucleare, Sezione di Roma,			   
University di Roma I, ``La Sapienza," I-00185 Roma, Italy}\\					   
\rUP {48} {\eightit Rutgers University, Piscataway, New Jersey 08855} \\				   
\rUP {49} {\eightit Texas A\&M University, College Station, Texas 77843} \\			   
\rUP {50} {\eightit Texas Tech University, Lubbock, Texas 79409} \\				   
\rUP {51} {\eightit Istituto Nazionale di Fisica Nucleare, Universities of Trieste and       	   
Udine, Italy} \\										   
\rUP {52} {\eightit University of Tsukuba, Tsukuba, Ibaraki 305, Japan} \\			   
\rUP {53} {\eightit Tufts University, Medford, Massachusetts 02155} \\				   
\rUP {54} {\eightit Waseda University, Tokyo 169, Japan} \\					   
\rUP {55} {\eightit Wayne State University, Detroit, Michigan  48201} \\				   
\rUP {56} {\eightit University of Wisconsin, Madison, Wisconsin 53706} \\				   
\rUP {57} {\eightit Yale University, New Haven, Connecticut 06520} \\				   
\protect\end{center}
\protect\vskip .026in%
}
\noaffiliation


\date{December 4, 2003 
}

\begin{abstract}
  We report the observation of a narrow state
  decaying into $J/\psi\pi^+\pi^-$ and produced in $220~\ipb$ of $\bar{p} p $
  collisions at $\sqrt{s}=1.96~\TeV$ in the CDF~II experiment. 
  We observe   $ 730 \pm 90 $   decays.
  The mass is measured to be  
$3871.3 \pm 0.7\,(stat) \pm 0.4\,(syst)~\MeVcc$,
  with an observed width consistent with the detector resolution.
  This is in agreement with the recent observation by the Belle Collaboration 
  of the $X(3872)$ meson. 
\end{abstract}


\maketitle


The study of bound states of charm-anticharm quarks
revolutionized our understanding of hadrons
beginning with the discovery of the $J/\psi$ meson in 1974~\cite{JPsiDiscov}.
Although numerous charmonium
($c\bar{c}$) states are now known, others should be observable.
Recently, the Belle Collaboration reported a new particle, $X(3872)$,
observed in exclusive decays of $B$ mesons produced in $e^+e^-$
collisions~\cite{Belle3872}.  This particle has a mass of 3872~\MeVcc\ and
decays into $J/\psi\pi^+\pi^-$. A natural
interpretation of this particle would be a previously unobserved
charmonium state, but there are no such states  predicted to lie at
or near the observed mass with the right quantum
numbers to decay into $J/\psi\pi^+\pi^-$~\cite{CharmSpec, CharmHypoth}.
Within the framework of QCD there
are other possibilities~\cite{AltHypoth}. 
The near equality of
the $X(3872)$ mass to the sum of the $D^0$ and ${D}{^{*0}}$  masses
suggests that the $X(3872)$ may be a deuteron-like ``molecule'' composed of a
$D$  and $\overline{D}{^{*}}$. Another possibility is that the $X(3872)$ is
a $c\bar{c}g$ hybrid meson---a $c\bar{c}$ system possessing a valence gluon.
These novel possibilities have excited great interest
in the $X(3872)$~\cite{Excite}. Whether it is a new form of hadronic matter
or a conventional $c\bar{c}$-state in conflict with theoretical
models, the $X(3872)$ is an important object of study.
Here we report the  observation of a  $J/\psi\pi^+\pi^-$ resonance produced 
inclusively in $\bar{p}p$ collisions and which is consistent with the $X(3872)$.

The analysis uses a data sample of $\bar{p}p$ collisions at $\sqrt{s}=1.96~\TeV$
 with an integrated luminosity of $220~\ipb$  collected between February 2002 and
August 2003  with the upgraded Collider Detector (CDF~II)
at the Fermilab Tevatron.
The important components of the CDF~II detector for this analysis include a tracking
system composed of a silicon strip vertex detector (SVX~II)~\cite{SVXII}
surrounded by an open cell drift chamber system called the Central Outer Tracker
 (COT)~\cite{COT}.   
The SVX~II detector comprises five concentric layers of double-sided sensors
located at radii between 2.5 and 10.6 cm.
On one side of the sensors axial strips
measure positions in the plane transverse to the beamline.
Strips on the other side are used for stereo measurements.
One layer has strips tilted by  $+1.2^{\circ}$, another by 
 $-1.2^{\circ}$,  and three layers by  $90^{\circ}$
with respect to the axial strips.
The active volume of the COT is a 3.1 m long cylinder covering radii from 43 to 132~cm 
with 8~superlayers of 12~wires each.
Superlayers of axial wires alternate with superlayers of $+2^{\circ}$
 stereo angle wires and superlayers of $-2^{\circ}$ stereo angle wires 
to provide three-dimensional tracking.  The  central tracking
system is immersed in a 1.4 T solenoidal magnetic field for the measurement of
charged particle momenta transverse to the beamline, $p_T$.
The outermost detection system consists of planes of multi-layer drift 
chambers for detecting muons~\cite{Muons}.
The Central Muon system (CMU) covers
 $\left| \eta \right| \leq  0.6$,
where $\eta \equiv -\ln[\tan(\theta/2)]$ and $\theta$ is the angle 
of the particle with respect to the direction of the proton beam.  
Additional muon chambers (CMX)  extend the rapidity coverage to
$\left| \eta \right| = 1.0$.

In this analysis  $J/\psi\to\mu^+\mu^-$ decays are recorded using a dimuon
trigger. The CDF~II detector has a three-level trigger system.  The Level-1
trigger uses tracks in the muon chambers with a clear separation 
in azimuth from neighboring tracks.
 The eXtremely Fast Tracker
(XFT)~\cite{XFT} uses information from the COT to select tracks based on
$p_{T}$.  XFT tracks with 
$p_T \geq 1.5~\GeVc$ ($p_T \geq 2.0~\GeVc$) are
extrapolated into the CMU (CMX) muon chambers and compared with the positions of
muon tracks.  If there are two or more XFT tracks with matches to muon
tracks, the event passes the Level-1 trigger.  
Dimuon triggers have no requirements at Level 2.
At Level 3, the full tracking information from the COT
is used to reconstruct a pair of opposite sign muon candidates
in the mass range from $2.7$ to $4.0~\GeVcc$.  
Events passing the Level-3 trigger are recorded for further analysis.

The offline analysis makes use of the best available calibrations of the
tracking system for reconstructing events. 
Well reconstructed tracks are selected by accepting only those with
$\ge 3$ axial SVX~II hits, and  $> 20$ axial and  $> 16$ stereo COT hits.
Tracks are refit to take into account the ionization energy loss 
appropriate for the particle hypotheses under consideration~\cite{Dmass}.  
Dimuon candidates are selected in the mass range from
$2.8$ to $3.2~\GeVcc$ after being constrained to originate from a common point
in a three-dimensional vertex fit. 
Pairs of oppositely charged tracks,
both having $p_T \geq 0.35~\GeVc$ and assumed to be pions,
are then fit with the dimuon candidates
to a common vertex.
In this three-dimensional vertex fit the dimuon mass is
constrained to be the world average $J/\psi$ mass~\cite{PDG}.
We require that the $\chi^2$ for the $J/\psi \pi^{+} \pi^{-}$ vertex fit
must be less than $40$.

Due to the large multiplicity of charged tracks in some events,
there can be a large number of $J/\psi \pi^+ \pi^-$ candidates in one event
that  satisfy the pre-selection cuts discussed above,
especially for larger  $J/\psi \pi^+ \pi^-$ masses. 
These events contribute a large amount of combinatoric background
relative to a small potential signal.
We reject events with  12 or more pre-selection candidates
that have masses below 4.5 $\GeVcc$. 
Although a large number of candidates are accepted at this stage,
after the final selection the average number of  $J/\psi \pi^+ \pi^-$ 
candidates within the mass window of interest  (3.65-4.0~\GeVcc)
is fewer than $1.2$ per event
for events with at least one such candidate.
The specific number of pre-selection candidates allowed per event 
is determined  by the optimization procedure described below.

We apply the  following tighter cuts to suppress $J/\psi \pi^{+} \pi^{-}$ backgrounds:
$\chi^2<15$ for the  dimuon vertex fit,
dimuon invariant mass within $60~\MeVcc$ 
($\sim\!4$ standard deviations) of the world average $J/\psi$ mass,
$p_T(J/\psi) \geq 4~\GeVc$,  
$\chi^2<25$  for the  $J/\psi \pi^{+} \pi^{-}$ vertex fit, 
$p_T(\pi) \geq 0.4~\GeVc$,
and $\Delta R \leq 0.7$ for both pions.  
Here $\Delta R$ is defined as $\sqrt{ (\Delta \phi)^{2} + (\Delta
\eta)^{2} }$ where $\Delta \phi$ and $\Delta \eta$ are the azimuthal angle 
and pseudorapidity of the pion with respect to 
the $J/\psi \pi^{+} \pi^{-}$ candidate.

The values of these cuts are determined by an iterative optimization 
procedure in which the significance $S/\sqrt{S+B}$ is maximized, 
where $S$ and $B$ respectively represent the numbers 
of signal and background candidates.
$B$ is obtained from a background fit to the data in a window
around $3872~\MeVcc$.
The dependence of the $X$-yield on the cuts is modeled by using 
the observed $\psi(2S)$ signal. 
The value used for $S$ is obtained by rescaling the $\psi(2S)$-yield
to reflect the much smaller $X(3872)$ signal.
The rescaling factor is determined such that $S$ 
matches the observed $X$-yield for a set of reference cuts.
Since the denominator of the significance ratio is dominated by the much
larger background the optimization is not sensitive to the precise
value of the rescaling.

\begin{figure}
\includegraphics[scale=0.42]{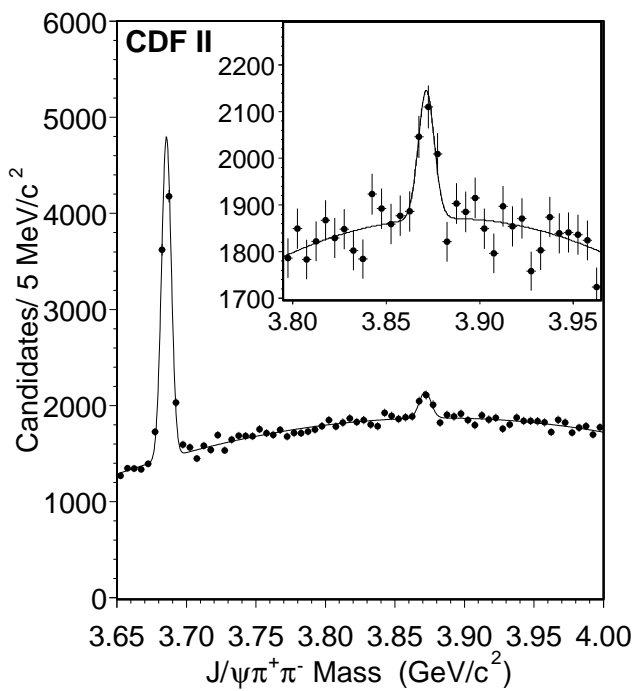} 
\caption{\label{cut1} 
The mass distribution of $J/\psi\pi^{+}\pi^{-}$ candidates
passing the selection described in the text. A large peak for the $\psi (2S)$
 is seen and a  signal near a mass of $3872~\MeVcc$ 
is visible (enlargement shown in the inset).
The curve is a fit using two Gaussians and a
quadratic background to describe the data.}
\end{figure}

\begin{figure}
\includegraphics[scale=0.42]{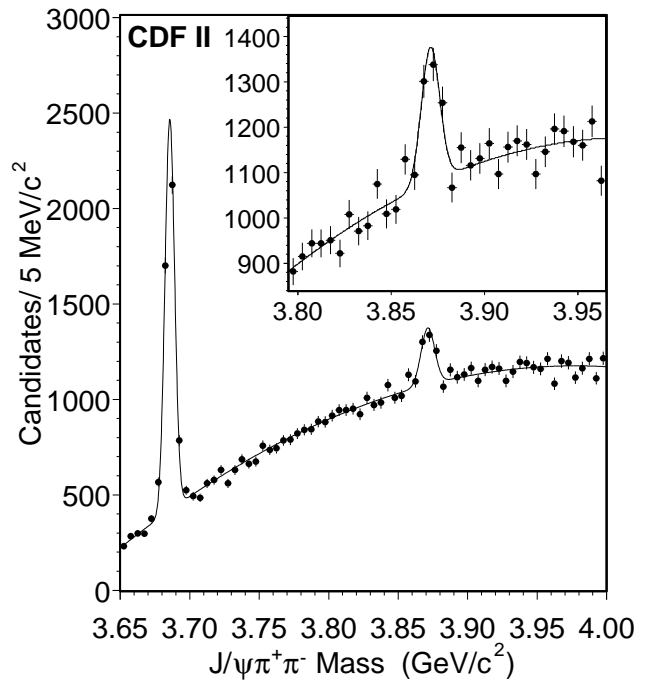} 
\caption{\label{cut2} 
The mass distribution of $J/\psi\pi^+\pi^-$ candidates requiring 
 $m(\pi^+\pi^-)>500~\MeVcc$. The curve is a fit with two
 Gaussians and a quadratic background.}
\end{figure}

The  $J/\psi \pi^{+} \pi^{-}$ mass distribution of the selected candidates 
is displayed in Figure~\ref{cut1}.
A large peak for the $\psi (2S)$ is seen, and in addition, a small peak at a 
$J/\psi \pi^{+}\pi^{-}$  mass around $3872~\MeVcc $ is observed. 
To fit the mass distribution, we model each peak by a single Gaussian 
and use  a quadratic polynomial to describe the background.
A  binned maximum likelihood  fit of the mass spectrum 
between $3.65$ and $4.0~\GeVcc$ is also shown in  Figure~\ref{cut1}.
The fit yields signals of 
$5790 \pm 140$ $\psi(2S)$ 
candidates and
 $580 \pm 100$  $X(3872)$ 
candidates.

The $X(3872)$ signal reported by the Belle Collaboration
favors large $\pi^{+}\pi^{-}$ masses. Our data support this conclusion as well. 
Figure~\ref{cut2} shows the $J/\psi \pi^{+} \pi^{-}$ mass distribution 
after requiring the  $\pi^{+}\pi^{-}$ invariant mass to be above $500~\MeVcc$,
a value large enough to probe the high mass behavior of the  $X(3872)$ candidates
and yet not eliminate all the $\psi(2S)$ reference signal. 
Fitting the mass spectrum between $3.65$ and $4.0~\GeVcc$  gives  
 $3530 \pm 100$ $\psi(2S)$ 
candidates and
 $730 \pm 90$  $X(3872)$ 
candidates. 
The fitted mass and width of the
$\psi(2S)$ are 
$3685.65 \pm 0.09\,(stat)~\MeVcc$ 
and 
$3.44 \pm 0.09\,(stat)~\MeVcc$,
respectively. 
For the  $X(3872)$ we obtain a mass of 
$3871.3 \pm 0.7\,(stat)~\MeVcc$
and a width of $4.9\pm0.7~\MeVcc$.
The latter value is consistent with detector resolution.
Our mass is in good agreement with the Belle result of  
$3872.0\pm0.6\,(stat)\pm 0.5\,(syst)~\MeVcc$~\cite{Belle3872}.

Imposing the dipion mass cut reduces the background 
by almost a factor of two, and apparently increases the
amount of fitted $X(3872)$ signal.
A significant part of the increase is
attributable to a larger fitted width.
The original fit without the $500~\MeVcc$ cut returns a smaller
but consistent width of $4.2\pm0.8~\MeVcc$.
We conclude that the  $X(3872)$ signal yield after the dipion
cut is unchanged within statistics,
and thus there is little signal with dipion 
masses below $500~\MeVcc$.
We use the selection with the dipion mass cut for measuring
the  $X(3872)$ mass as the improved signal-to-noise ratio 
reduces the statistical  uncertainty.

 The fit displayed in Figure~\ref{cut2}  has a $\chi^{2}$ of 74.9 for 61 degrees of
freedom, which corresponds to a probability of 
$10.9\%$. 
To estimate the significance of the signal, we first count the number 
of candidates in the 3 bins centered on the peak, 
i.e.~3893. 
The 3-bin background is estimated from the fit to be 
3234 
candidates, leaving a signal of 
$659$ 
candidates. 
In a Gaussian approach, this corresponds to a  significance of 
$659/\sqrt{3234}=11.6$ 
standard deviations.
The Poisson probability for 
3234
to fluctuate up to or above 
3893
is in good agreement with the Gaussian estimate,
considering the approximations of each method.

The systematic uncertainty on the mass scale is related to the momentum scale
calibration, the various tracking systematics, and the vertex fitting.  These
effects were studied in detail for our measurement of the mass difference
$m(D^{+}_{s}) - m(D^{+})$~\cite{Dmass}, where the systematic uncertainty was 
$\pm0.21~\MeVcc$. A larger systematic uncertainty arises for our $X(3872)$ mass
determination because it is an absolute measurement.  
We use the $\psi(2S)$ mass to gauge our systematic uncertainty.
With the dipion mass cut, the  $\psi(2S)$ mass 
is measured to be $0.3~\MeVcc$ below the world average mass of 
$3685.96\pm0.09$~\cite{PDG},
a difference substantially larger than the statistical uncertainty of $0.1~\MeVcc$.
Studies of the stability of the  $\psi(2S)$ mass 
for different selection requirements indicate a slightly larger systematic 
uncertainty of $0.4~\MeVcc$ should be assigned. 
Variations of the fit model and fit range have negligible effect on the mass.

In summary, we report the observation of a state consistent with
the $X(3872)$ decaying into $J/\psi \pi^{+} \pi^{-}$.
From a sample of 
 $730\pm90$ 
candidates we measure the $X(3872)$ mass to be 
$3871.3\pm0.7\,(stat)\pm0.4\,(syst)~\MeVcc$,
and find that the observed width is consistent with the detector resolution.
This is in agreement with the measurement by the Belle 
Collaboration using $B^\pm$ decays~\cite{Belle3872}.
The average mass from the two experiments, 
assuming uncorrelated systematic uncertainties, is 
$3871.7 \pm0.6~\MeVcc$.
Our large sample of  this new particle opens up avenues
for future investigations, such as production mechanisms, 
the dipion mass distribution, and  spin-parity analysis.

We thank the Fermilab staff and the technical staffs of the participating institutions 
for their vital contributions. This work was supported by the U.S.
Department of Energy and National Science Foundation; 
the Italian Istituto Nazionale di Fisica Nucleare; the Ministry of Education, Culture, Sports,
Science and Technology of Japan; the Natural Sciences and Engineering Research Council of Canada; 
the National Science Council of the Republic
of China; the Swiss National Science Foundation; the A.P. Sloan Foundation; 
the Bundesministerium f\"ur Bildung und Forschung, Germany; the
Korean Science and Engineering Foundation and the Korean Research Foundation; 
the Particle Physics and Astronomy Research Council and the
Royal Society, UK; the Russian Foundation for Basic Research; 
the Comisi\'on Intermi\-nis\-terial de Ciencia y Tecnolog\'{\i}a, Spain; 
the European Community's Human Potential Programme under contract HPRN-CT-20002, 
Probe for New Physics; and 
the Research Fund of Istanbul University Project No. 1755/21122001.


\end{document}